\documentclass[conference]{IEEEtran}

\usepackage{graphicx}
\usepackage{amsmath}
\usepackage{url}
\usepackage[dvipsnames]{xcolor}
\usepackage[justification=centering]{caption}
\usepackage{subcaption}

\DeclareMathOperator*{\argmax}{arg\,max}

\ifCLASSINFOpdf
\else
\fi

\begin{document}
%
\title{The Predictive Power of Social Media: On the Predictability of U.S. Presidential Elections using Twitter}

\author{\IEEEauthorblockN{Kazem Jahanbakhsh}
\IEEEauthorblockA{Computer Science, PhD\\
Vancouver, British Columbia\\
Email: k.jahanbakhsh@gmail.com}
\and
\IEEEauthorblockN{Yumi Moon}
\IEEEauthorblockA{Sociology Department, Yonsei University\\
Seoul, South Korea\\
Email: yumimoon@yonsei.ac.kr}}


%


\maketitle

\begin{abstract}
Twitter as a new form of social media potentially contains useful information that opens new opportunities for content analysis on tweets. This paper examines the predictive power of Twitter regarding the US presidential election of 2012. For this study, we analyzed 32 million tweets regarding the US presidential election by employing a combination of machine learning techniques. We devised an advanced classifier for sentiment analysis in order to increase the accuracy of Twitter content analysis. We carried out our analysis by comparing Twitter results with traditional opinion polls. In addition, we used the Latent Dirichlet Allocation model to extract the underlying topical structure from the selected tweets. Our results show that we can determine the popularity of candidates by running sentiment analysis. We can also uncover candidates popularities in the US states by running the sentiment analysis algorithm on geo-tagged tweets. To the best of our knowledge, no previous work in the field has presented a systematic analysis of a considerable number of tweets employing a combination of analysis techniques by which we conducted this study. Thus, our results aptly suggest that Twitter as a well-known social medium is a valid source in predicting future events such as elections. This implies that understanding public opinions and trends via social media in turn allows us to propose a cost- and time-effective way not only for spreading and sharing information, but also for predicting future events. 
\end{abstract}



%
\IEEEpeerreviewmaketitle

\section{Introduction}
An online-based community has emerged because of the rapid growth of the Internet usage, which has led to the possibility of horizontal and free communication \cite{castle:2000}. Seventy five percent of the Internet surfers used social media in the second quarter of 2008, which is a considerable growth from fifty six percent in 2007 \cite{kaplan:2010}. In addition, the number of cellphone users exceeded 3.3 billions in 2009, and smartphones technology has been embedded into people's everyday lives, providing location-aware services and empowering users to generate and to access multimedia contents \cite{demirbas:2010}. This technology breakthrough has affected social media to become ubiquitous and indispensable \cite{asur:2010}. 

Kaplan and Haenlein (2010) differentiated the concept of social media from Web 2.0 and user-generated content, explaining that the growth of high-speed internet access brought about the creation of social networking sites, which in turn coined the term social media that they categorized by characteristics including collaborative projects, blogs, content communities, social networking sites, virtual game worlds, and virtual social worlds. Even earlier than Facebook's launch in 2004, social media started its formation. For instance, Sixdegrees started its service in 1997 via which users could create their profiles, list and add friends \cite{kietzmann:2011}.

Social media has revolutionized the conventional media outlets (i.e. radio, television, newspapers and books) which have limits bounding up in times and in spaces \cite{bride:2003}. For instance, the appearance of social media services such as Facebook, Twitter, Flickr, Instagram, and numerous others has restructured the way in which information and news, such as Hudson river accident in 2009 and Arab Spring in 2010, spread in the world \cite{kazem:2012}. Furthermore, social media has become a multifunctional platform, giving users the opportunity to express, share, and influence, replacing traditional forms of media. Social media has transformed the Internet landscape from a simple platform for information itself to a dominant platform as a tool for influence, such as political influence \cite{hanna:2011}. This has contributed to the restructuring of free communicative space and to revision of the understanding of the public sphere conventionally conceived \cite{keane:1995}. Importantly, in facing the era of social media, social media data has been used as a source for predicting future events, such as political elections, in a short period of time without cost for collecting social media data as contrasted with traditional ways of analyzing public opinions using public polls. 

However, some researchers argue that social media is not as reliable as some have evaluated it to be, nor credible as a source to anticipate future events. A possible reason may be that social media is a virtual sphere, dominated mainly by computer holders \cite{papachrissi:2002}, and by those who are relatively comfortable with utilizing technology, such as wireless technologies and smartphones \cite{kazem:2012}. This may leave less credibility to social media to be a representative sample of the entire population. Having said that, social media itself is crucial since it allows people to engage in a horizontal civil society beyond limits that a physical society has, and since social media contents can be used as a source to anticipate real-world outcomes \cite{asur:2010}. Analyzing social media data in order to predict the future is appealing and useful for economic, medical, political, and social purposes. 

Before formulating our prediction framework, we needed to find an event with measurable output such as stock values in the future or political elections outcomes. We also needed to make sure that the event has high socio-economic value, and there are data available for prediction. We chose the US 2012 presidential election as the event to predict and Twitter as the source for data. US presidential election generated a large number of conversations in Twitter, and it satisfies the criteria listed above. We analyzed Twitter data to extract certain trends for the US election such as polarity trends. To predict the election's outcome, we collected tweets from September 29, 2012 until November 16, 2012 and selected political-related tweets only. We analyzed the political tweets to examine if there were any interesting patterns or trends, and if we could predict the results of the US election. 

To the best of our knowledge, very few studies have been devoted to employ a combination of different machine learning and natural language processing models to analyze a very large number of political tweets. None has carried out a deep comparison of predicting election using Twitter data with traditional polls either. In this work, we performed content analysis by running sentiment analysis and topic modeling on a representative sample of tweets within a three-month time span, and we compared our sentiment analysis results of Twitter with traditional pollster results within the same time span. We specifically tried to answer the following research questions:

\begin{itemize}
\item Can one use Twitter data to predict the 2012 US presidential election?
\item Are the content analysis results of Twitter comparable with traditional pollster results? 
\item Can one use topic modeling in order to discover topics of Twitter-based discussions? Are those extracted topics in match with offline discussions?
\end{itemize} 

Some of our major findings are as follows: (1) Twitter can be used as a powerful source of information to predict elections results such as the 2012 US election. (2) Our prediction results from political tweets are in match with pollster results, especially when we get closer to the election day; however, using Twitter data for prediction allows us to get real-time insights from people's opinions in a virtual space. (3) Geographical sentiment analysis of tweets allows us to predict the outcomes of the US election for $76\%$ of states successfully. (4) By using topic modeling, one can discover hot political topics discussed in social media. These findings can potentially benefit political experts who run political campaigns.

The remainder of this paper is organized as follows: Section \ref{section:relatedwork} reviews the recent work in the field. Section \ref{section:problem} defines the problem to be tackled. Section \ref{section:data} explains how we collected political tweets and describes the properties of collected data. Section \ref{section:analysis} describes the machine learning and language processing techniques that we used to analyze the distributions of political tweets and to examine the predictive power of Twitter data for forecasting the election results. Finally, Section \ref{section:conclusion} concludes the paper.

\section{Related Work}
\label{section:relatedwork}
A social medium Twitter service has become omnipresent with hyper-connectivity for social networking and content sharing. Facebook established a status update field in June 2006, but Twitter took status sharing between people to cell phones four months later \cite{demirbas:2010}. Since then, Twitter has grown exponentially beyond status sharing. For instance, in 2009 the number of Twitter users was 18.2 million, which was 1,448 per cent growth from the year of 2008 \cite{marwick:2010}. Then, what made Twitter grow dramatically? One reason may be the simplicity of Twitter in using. For instance, while blogging needs decent writing skills and a large size of content to fill pages \cite{demirbas:2010}, Twitter originally developed for mobile phones \cite{marwick:2010} restricts users to posting 140-character text messages, also known as tweets, to a network of others without technical requirement of reciprocity \cite{castillo:2011}. This encourages more users to post, and which facilitates real-time diffusion of information \cite{castillo:2011}. Thus, users can easily post and read tweets on the web, using different access methods, such as desktop computers, smartphones, and other devices \cite{marwick:2010}. 

According to Pew Research, the percentage of the Internet users who are on Twitter currently stands at 16 percent. Individuals under 50 and in particular those in the age range of 18 to 19 are most likely to use Twitter. In addition, urban dwellers are more likely to be on Twitter than both suburban and rural residents \cite{duggan:2013}. For instance, Mislove et al. found that Twitter users in the US are significantly overrepresented in populous states and are underrepresented in much of mid-west since the Twitter representation rate increases as the population of a state increases \cite{mislove:2011}. Some studies point out the drawback of Twitter demographic profiles, since it is relatively difficult to calculate demographics of Twitter users and to detect users' ages because information about those users' self-reports is limited \cite{mislove:2011}. Although numerous users share their identities on social media sites, many others do not open their social profiles, and some users present even made-up identities intentionally because of the concerns that their information could be used as a source for data mining and surveillance \cite{kietzmann:2011}. This results in the lack of knowledge about users' demographic profiles, the sets of users' characteristics such as nationality, spoken language, and affiliation \cite{kazem:2012}. 

Marwick argued that Facebook or Twitter users' imagined audience might be different from actual readers who would be interested in tweets and posts \cite{marwick:2010}. Another account may lie in a big data fallacy that can be found in demographic bias in that users tend to be young, and big data itself may not be statistically representative of the whole population. For instance, Twitter users are predominantly males, but the rate of male users was found to have decreased, oppositely to the bias that the rate of male users would increase \cite{mislove:2011}. Because of these reasons, some contend that the predictability of Twitter data does not guarantee generalization of its positive analysis results in the past \cite{gayo_poll:2011}. Having a specific conception of the users' online identity presentation, and understanding of Twitter users can help for advanced observations and predictions, since such an understanding can reduce biases of tweet-based analysis \cite{mislove:2011}. While conducting this study, we could partly detect users' gender and resident cities.

There is a claim that there is no correlations between social media data and electoral predictions since Twitter data anlayzed by using lexicon-based sentiment analysis did not predict the 2010 US congressional elections \cite{gayo_limits:2011}. Likewise, Google Trends was not predictive for the 2008 and the 2010 US elections \cite{lui:2011}. However, their claims may not generalize since it is based on comparing the candidates who won the 2008 congressional elections to the candidates whose names were searched frequently on Google or appeared on Twitter. In order to obtain more credible results, more systematic methods of detecting would be necessary whether or not candidates names were used positively, negatively, or neutrally, instead of focusing on the frequency rates of the candidates' names who simply were searched on Google or appeared on Twitter. In addition, the tweets simply mentioning political parties are not sufficient either \cite{sang:2011}. Other reasons that cause prediction failure are inadequate demographic data, existence of spammers, propagandists, and fake accounts in social media \cite{gayo_limits:2011}.

In terms of methodology, we can say that our methods are systematic and credible. Regarding Gayo-Avello et al.'s claim \cite{gayo_limits:2011} that current methods using sentiment analysis on Twitter data for predicting the results of elections are not better than random classifiers, we think that the number of tweets (234,697) they collected is not sufficient to be representative of the number of actual voters. In addition, their sentiment analysis on 2800 words including 2,325 tweets (positive, negative, and neutral) is not a large sample size compared to our Twitter data. We agree with the claim that it is not trivial to identify likely voters via social media, but disagree with the argument that it is hard to obtain unbiased sampling from likely voters in social media \cite{gayo_limits:2011}. In contrast with the argument, our Twitter data suggests that it is possible to collect an unbiased and random sample from those users who at least mention about a specific election in social media sites.

In the meanwhile, a large number of people have raised their voices via social media, which led to the emergence of companies that provide prediction services by using social media data. Topsy can be an example of an analysis company using social media data, providing a technology service for analyzing conversations in social media sites such as Twitter and Goggle+, and drawing insights from the conversations \cite{topsy:2014}. Intrade.com also provides a prediction service by utilizing a market trading exchange model predicting the probability of an event to occur in the future \cite{intrade:2014}. For the 2012 US presidential election, Intrade market accurately predicted the outcomes of all the US state electoral contests except Florida and Virginia. Although there are some claims against Google Trends we mentioned earlier, Google Trends is still useful technology in some degree, by which people can search keywords on Google website in order to detect current issues and trends\cite{googletrend:2014}. Searching keywords is often correlated with various economic indicators and may help for short-term economic prediction \cite{choi:2011}. 

Surprisingly, based on the analysis of 50 million tweets they collected, Ritterman et al. proved that even noisy information in social media such as Twitter can be used as a proxy for public opinions \cite{ritterman:2009}. They found that Twitter data provides more than factual information about public opinions on a specific topic, yielding better results than information from the prediction market. For instance, Twitter data can be used as social sensors for real-time events and to forecast the box-office revenues for movies \cite{asur:2010}. The rate at which tweets posted about a particular movie has a strong positive correlation with the box-office gross. In addition, Twitter as a platform can reflect offline political sentiment validly \cite{tumasjan:2010}, and it can mirror consumer confidence. Interestingly, even a mere number of Twitter data dominated by a small number of heavy users predicted an election result, such as the results of the 2011 Irish general election \cite{bermingham:2011}, and even came close to traditional  election polls \cite{tumasjan:2010}. This aptly suggests that social media data, especially Twitter, can replace the costly- and time-intensive polling methods \cite{connor:2010}. Thus, we argue that social media can be used as a credible source for predicting the near future.  

\section{Problem Definition}
\label{section:problem}
In this paper, we analyzed political tweets in order to find interesting trends and patterns which allow us to predict the US presidential election of 2012. To be formal, we analyzed a set of tweets in a given time interval $[t_{s},t_{e}]$ where $t_{s}$ (i.e ``2012-09-29 21:20:18 UTC'') and $t_{e}$ (i.e. ``2012-11-16 14:32:29 UTC'') denote the timestamps for the first and last observed tweets, respectively. We use $T$ to denote the set of observed tweets within this time frame where each tweet $tw \in T$ is represented by a set of attributes including: \textit{unix timestamp}, \textit{source}, \textit{author}, \textit{latitude}, \textit{longitude}, and \textit{text}. 

The first set of problems that we address in this paper is to compute temporal probability distributions of different properties of tweets in $T$. In particular, we would like to compute the distribution for the number of posted tweets per day in PST timezone. Next, for each day prior to the election, for each specific word $w$ (e.g. \textit{Obama}), we compute the number of tweets containing that word. We proceed our analysis by extracting distributions of frequent hashtags in posted tweets per day before the election.

For a given tweet $tw$, we can compute a polarity likelihood $P(l|tw)$ where $l$ denotes the polarity of tweet $tw$ and can take any values from label set $L=\{0, -1, +1\}$. We label tweet $tw$ with label $l$ which produces the maximum  polarity likelihood. We run our first sentiment analysis by computing the polarity likelihoods of a subset of tweets randomly sampled from each day prior to the elction day. We stduy distributions for the number of \textit{positive} and \textit{negative} tweets for each candidate per day. Second, we run a geo-spatial version of our sentiment analysis where we compute the popularity of each candidate in each US state by sampling only those tweets with location information.  

Finally, as the last problem we randomly sample a subset of tweets from each day prior to the election. Next, we compute the probability distributions for unobserved topics $P(w_{i}|topic\_index=k)$ where $w_{i} \in V$ is a word from tweets vocabulary and $k$ is topic index. Our goal here is to extract unobserved topics discussed in Twitter by mining a large random sample of tweets per day.  

\section{Data Collection}
\label{section:data}
Predicting future events using big data has been in the core of attention in the last few years. To tackle this problem, we needed to find an event that we could measurably predict and to find big data to drive the prediction. We found that predicting the US presidential election was an appealing problem. We also found reports, which have shown that both Republican and Democratic campaigns spent considerable amount of time to promote their candidate in social media such as Twitter and Facebook. In particular, we have seen several reports highlighting Obama's efforts on exploiting social media for the 2008 and 2012 US presidential elections \cite{obama:2014,nytimes:2008}. Thus, we decided to collect data from Twitter website for analyzing tweets in order to predict the US presidential election. Choosing Twitter was mainly because of its open API which made collecting stream of tweets be convenient.

We implemented a Twitter crawler that uses Twitter Streaming API to collect political-related tweets in real-time \cite{twitter:2014}. We developed a simple listener in Python for collecting political tweets using Tweepy Python library \cite{tweepy:2014}. We filtered twitter stream in order to collect political tweets only related to the 2012 US presidential election. For such, we filtered tweets that contain political keywords such as \textit{barack obama}, \textit{mitt romney}, \textit{us election}, \textit{paul ryan}, \textit{joe biden}, and so on.  We ran our Python crawler from September 29, 2012 until November 16, 2012, by then we collected around 39 million tweets. The number of collected tweets before the election day (i.e. November 6, 2012) was around 32 million tweets among which 140 thousand tweets came with geo-locations (i.e. $0.43\%$). We stored all tweets in a \textit{MySql} database. We also stored geo-tagged tweets in a \textit{k-d tree} data structure for running machine learning and data mining analysis. We added necessary indexes on the \textit{tweet} table in order to minimize running times of \textit{MySql} queries.

\begin{table}
   \centering
	\begin{tabular}{|c|}
	\hline
		RT @JCULLI: If Romney win Iâm moving to Canada. No bullshit.\\ \hline
		Mitt Romney buys followers. LO\\ \hline
		Samuel L. Jackson To Voters: âWake The F*ck Upâ And Vote Obama\\ \hline
		My ex bf is soo stupid routing 4 romney.\\ \hline
		Goin to vote. \#Romney\\ \hline
		\#Benghazi - \#Obama Linked To Benghazi Attac\\ \hline
		Hell yeah!!!!! \#Obama\\ \hline
	\end{tabular}
	\caption{A Sample of Political Tweets}
	\label{tab:tweets}
\end{table}

\begin{table}
   \centering
	\begin{tabular}{| c | c |}
	\hline
	    \textbf{Tweet Source} & \textbf{Freq Percentage}\\ \hline
		Twitter for iPhone & 32\%\\ \hline
		web & 22\%\\ \hline
		Twitter for Android & 20\%\\ \hline
		Echofon & 2.7\%\\ \hline
		Mobile Web & 2.5\%\\ \hline
		Twitter for iPad & 2.3\%\\ \hline
		TweetDeck & 2.3\%\\ \hline
		TweetCaster for Android & 1.8\%\\ \hline
		twitterfeed & 1.4\%\\ \hline
		Tweet Button & 1.1\%\\ \hline
	\end{tabular}
	\caption{Frequent Tweet Sources (Nov 6, 2012)}
	\label{tab:sources}
\end{table}

While collecting tweets, we made sure to collect a rich set of attributes associated with each tweet including: \textit{tweet content}, \textit{time of posted tweet}, \textit{author of tweet}, \textit{source}, and \textit{tweet location}. Table \ref{tab:tweets} shows a sample of tweets contents. As we see, political tweets are usually short and do not use a formal language. \textit{Source} attribute is the platform used by Twitter user for tweeting. Table \ref{tab:sources} shows the most frequent sources used for tweeting on election day (i.e. November 6, 2012). \textit{Tweet location} is composed of \textit{latitude} and \textit{longitude} where tweet was posted. The rich data we collected for the US election provides a unique opportunity to tap into public opinions. We have plans to make an anonymized version of our data publicly available in the near future.

\section{Mining Trends of Political Opinions}
\label{section:analysis}
Having access to 39 million tweets provides a unique opportunity to run deep statistical and machine learning analysis on tweets contents. In particular, having tweeting time information for posted tweets allows us to perform temporal statistical analysis on the political tweets. Moreover, having location data for a subset of our tweets enables us to perform time-spatial statistical analysis. In this section, we first run a basic statistical analysis on collected tweets in order to get insights about political trends in Twitter. Next, we design and implement an enhanced version of a machine learning classifier to compute sentiments of tweets. Finally, we run our in-house implementation of LDA algorithm by which we digg into the underlying discussions inside Twitter in the course of election.

\begin{figure}
\centering
	\includegraphics[width=\linewidth,height=5cm]{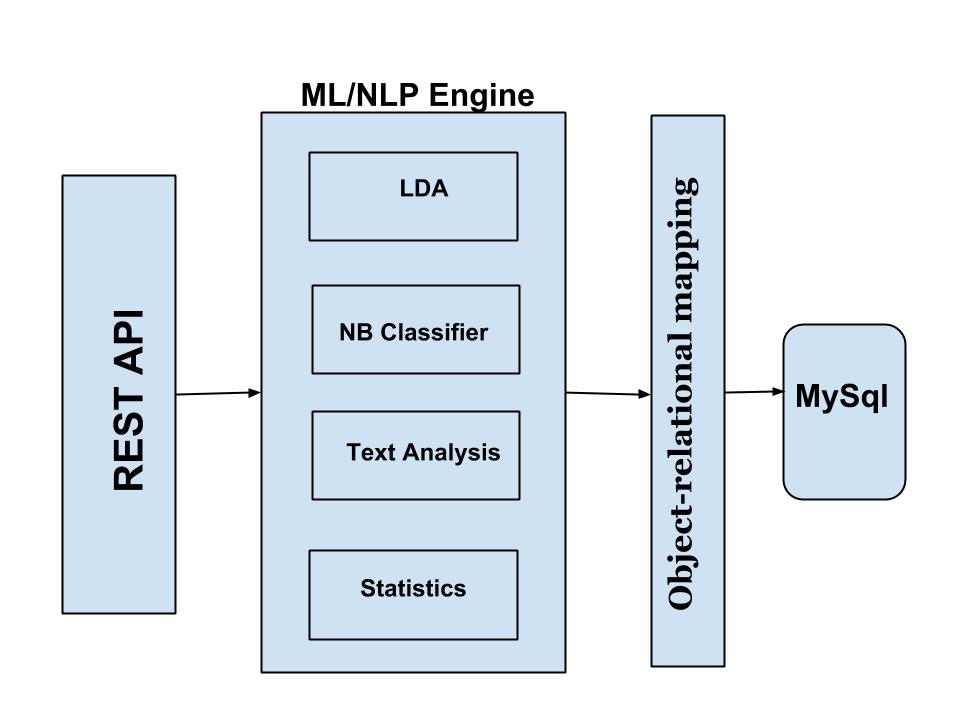}
	\caption{ML/NLP Software Architecture}
	\label{fig:ml_nlp_engine}
\end{figure}

One of our contributions in this paper is the development of an advanced machine learning and language processing (ML-NLP) engine which enables us to run text analysis on a large number of tweets. Figure \ref{fig:ml_nlp_engine} shows the complete software architecture of the ML-NLP engine which we have built in the last two years. We have implemented the ML-NLP software in the Java language. The ML-NLP engine accesses to the MySql database through the \textit{Hibernate ORM} which is an object-relational mapping library for the Java language \cite{hibernate:2014}. We also used \textit{Dropwizard} framework in order to put the ML-NLP engine behind a RESTful web service \cite{dropwizard:2014}. This allows researchers to run their analysis through REST API request without being worried about the implementation details. The ML-NLP engine is consisted of four major components: Statistical component which is responsible to compute all basic statistics, Text Analysis for running basic text analysis tasks, the Naive Bayes classifier for sentiment analysis, and the LDA algorithm for topic modleing. We have plan to make our ML-NLP software open source for the public use of other researchers in the field.

\subsection{Statistical Analysis}
We started our analysis by touching the surface of our big data. Our goal is to see if we can get insights about public opinions by running a few stat analysis on top of our tweets.

\subsubsection{Tweets Frequency Distribution}
\begin{figure}
\centering
	\includegraphics[width=\linewidth,height=8cm]{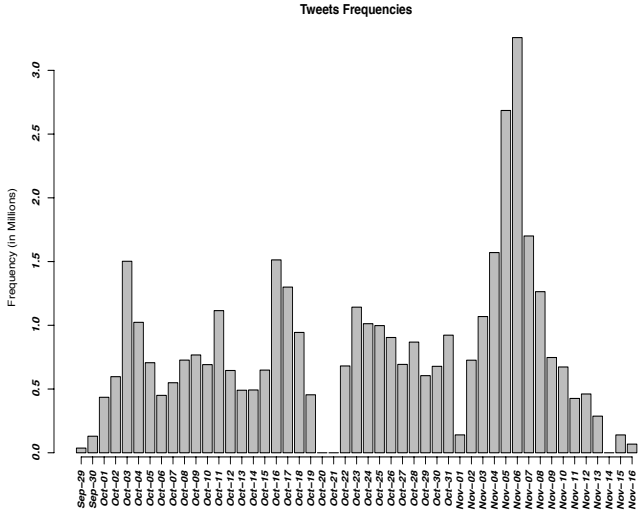}
	\caption{Political Tweets Frequency Distribution}
	\label{fig:freq}
\end{figure}

As the first analysis, we started with computing the tweets frequency distribution by which we  measured the number of political tweets posted on each day starting from September 29th and ending on November 16th, 2012. The results are shown in Figure \ref{fig:freq}. From the figure, we find local maximums occuring on October 3, October 11, October 16, October 22, and November 6, 2012. Table \ref{tab:debates} shows the schedule of important events for the 2012 US presidential election. As we see, there is a match between local maximums on Figure \ref{fig:freq} and important election events including the three debates and the election day. As we expected, the global maximum occured on the election day, which is over three million tweets. Readers should note that our crawler process crashed before third presidential debate due to some technical issues and was restarted on the same day.

\begin{table}
   \centering
	\begin{tabular}{|c|c|}
	\hline
	  \textbf{Date} & \textbf{Event}\\ \hline
	  Oct 3, 2012 & First presidential debate\\ \hline
	  Oct 11, 2012 & Vice-presidential debate\\ \hline
	  Oct 16, 2012 & Second presidential debate\\ \hline
	  Oct 22, 2012 & Third presidential debate\\ \hline
	  Nov 6, 2012 & Election day\\ \hline
	\end{tabular}
	\caption{US 2012 Presidential Election Schedule}
	\label{tab:debates}
\end{table}

\subsubsection{Tweets Mentions Distribution}
\begin{figure}
    \centering
    \includegraphics[width=\linewidth,height=8cm]{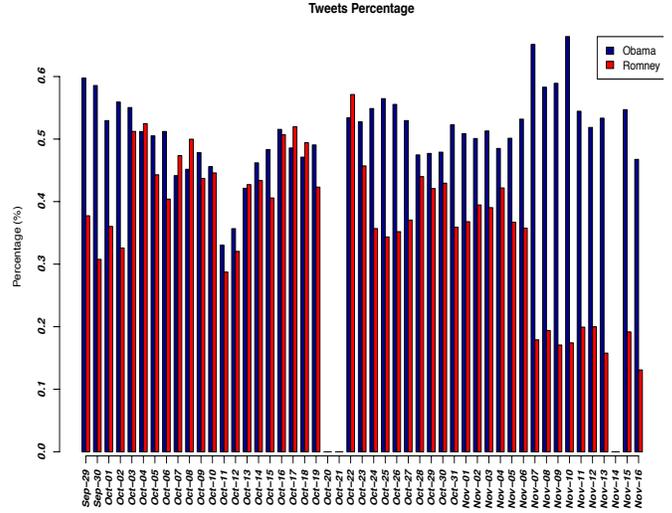}
    \caption{Obama and Romney Mentions Percentage in Twitter}
    \label{fig:mention}
\end{figure}

In the second analysis, we were interested in measuring how much each candidate was in the center of attention in Twitter. For such, we analyzed the tweets contents from September 29 until November 16. For each day, we computed the percentage of the tweets that mentioned either ``\textit{Romney}'' or ``\textit{Obama}''. To run the analysis, we randomly sampled 10K tweets from each day and computed the percentage of mentions for each candidate separately. The results of our analysis are shown in Figure \ref{fig:mention}. By analyzing the results closely, we make a few interesting observations. First of all, Obama mostly led Romney. In other words, Obama was more in the center of attentions in Twitter than Romney was. Interestingly, we observe that after each presidential debate, Romney led Obama for one or two days; however, after that, he lost the momentum again. This can be justified by the effort taken by Romney and his campaign to bring public attentions to him in social media space. Finally, from the third presidential debate onward, we observe that Obama by far led the Twitter social medium space. Thus, one can argue that after the last debate (i.e. October 22, 2012), the public opinions especially the undecided individuals were formed, which made Obama become the dominant player in Twitter discussions.

\subsubsection{Hashtags Distribution}
\begin{figure}
   \centering
   \includegraphics[width=\linewidth,height=8cm]{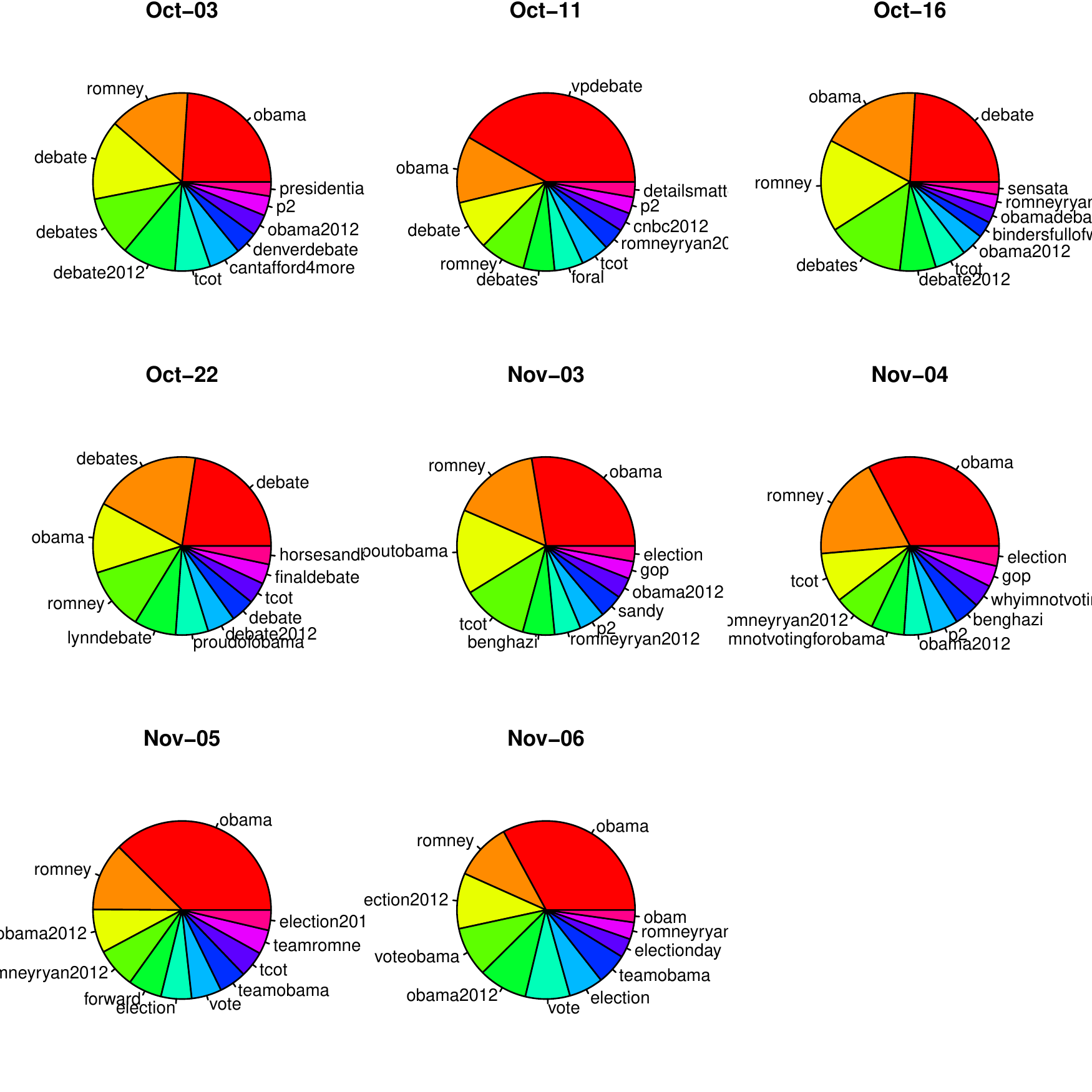}
    \caption{Distributions of Frequent Hashtags on Important Dates}
    \label{fig:hashtag}
\end{figure}

According to Twitter, hashtags preceded by ``\#'' symbol are used to mark important keywords or topics in a tweet. It was created organically by Twitter users to categorize messages. Therefore, by analyzing hashtags, one can discover popular topics in Twitter. Analyzing distributions of hashtags over time allows us to find out who is leading in Twitter and to measure popular topics that people discuss the most on the website. Figure \ref{fig:hashtag} demonstrates the hashtag distribution for eight different days including the debates, days leading to the election day, and the election day (i.e. November 6th, 2012). We can make a few interesting observations by analyzing the hashtags distributions. First, on the debate days, ``\#\textit{debate}'', ``\#\textit{obama}'', and ``\#\textit{romney}'' appeared as the most popular hashtags. Furthermore, as we approach the election day, ``\#\textit{obama}''  became the most popular hashtag in tweets. Thus, analyzing hashtags clearly demonstrates that Obama became the popular candidate in Twitter as the election day approached, which is in match with the result of the 2012 US presidential election.

\subsection{Sentiment Analysis}
As the second part of our analysis, we designed and developed an advanced version of \textit{Naive Bayes} classifier in order to compute sentiments of tweets \cite{pang:2005}. We chose the Naive Bayes classifier because it produces high quality results despite its simplicity. Below we explain how we designed, implemented and trained our NB classifier.

\subsubsection{Naive Bayes Classifier}
We are interested in computing the posterior probability of the label of a tweet given its content. Using Bayesian framework, we factorize posterior probability by product of prior probability (i.e. our prior belief) and likelihood, which is the knowledge we achieve from observational data. This relationship is shown in Equation \ref{bayes_equation}.

\begin{equation}
    \label{bayes_equation}
    Posterior \propto Prior \times Likelihood
\end{equation}

We make the independence assumption by which we assume that given the label of tweet, features (i.e. words) are independent random variables. Thus, we can factorize the likelihood probability $P(tweet|label)$ as the product of conditional probabilities of features given the corresponding label (i.e. $P(w_{i}|label=l_{j})$). This allows us to formulate our sentiment classification problem as shown in Equation \ref{nb_equation}.

\begin{equation}
    \label{nb_equation}
    P(label=l_{j}|tweet) \propto P(label=l_{j}) \times \prod_{i}P(w_{i}|label=l_{j}),
\end{equation}
where $P(label=l_{j}|tweet)$ is the posterior probability of tweet's label given its content, $tweet =\{w_{1}, ..., w_{n}\}$ and $l_{j} \in L=\{0, -1, +1\}$. Here, we model each tweet content as a bag of words. Also, each tweet can take any of three possible lables: $0$, $-1$, and $+1$ as its polarity indicating \textit{neutral}, \textit{negative}, and \textit{positive} sentiments, respectively. The label of a tweet is computed as the label with the maximum likelihood as shown in Equation \ref{classify_equation}.

\begin{equation}
    \label{classify_equation}
    label_{NB}=\argmax_{l_{j} \in L}P(label=l_{j}|w_{i}),
\end{equation}

To compute the likelihood probability $P(w_{i}|label=l_{j})$, we use the following smoothing equation:

\begin{equation}
    \label{indep_equation}
    P(w_{i}|label=l_{j})=\dfrac{count(w_{i}, l_{j}) + 1}{count(l_{j}) + |V|},
\end{equation}

where $count(w_{i}, l_{j})$ is number of times we have observed word $w_{i}$ with label $l_{j}$ in our training data, $count(l_{j})$ is the number of times we have seen label $l_{j}$ in our training data, and $|V|$ is the number of words in our training data (i.e. vocabulary size). 

For training the Naive Bayes classifier, we manually labeled \textbf{989} number of tweets. We selected these tweets uniformly at random from all of our tweets. Next, we went through the list of tweets and manually labeled each of them. For example, for ``\textit{RT @MarkSalling: Obama is going to win}'' we produced the following label ``\textit{obama=+1,romney=NA}'' meanining that the tweet only mentioned Obama with a positive polarity. As another example, we labeled ``\textit{Even after having gotten a raise this year, my lifestyle has been altered down because of the price of gas and food. Screw you Obama!}'' as ``\textit{obama=-1,romney=NA}''. Readers should note that for training NB classifier we used the tweets that mentioned only one of the two candidates. This is because when both candidates are mentioned in the same tweet, we needed to use a more sophisticated language processing technique in order to compute the right polarity for each candidate.

Before using a tweet for training or classification, we cleaned it and removed all noises. This includes removing \textit{URL}, ``@somebody'', ``RT'', ``\#something'', numbers, stop words, punctuations, and capitalization. Removing noise from tweets significantly improves the performance of the NB classifier. Next, we tokenized each tweet into its consisting words using \textit{space} as delimiter. We also modified our NB classifier to handle negation. In particular,  for the case that a negation word appeared in a tweet such as ``\textit{don't}'', we modified the tweet by adding \textit{NOT} keyword to the beginning on each word following the negation word until we see a punctuation. As an example, we change ``\textit{don't have favorite candidate, but ill vote for Obama anyway!}'' to ``\textit{don't NOT\_have NOT\_favorite NOT\_candidate, but ill vote for Obama anyway!''}. This modification allows us to handle negation successfully. Finally, we use \textit{logarithm} of probabilities while computing posterior probabilities $P(label=l_{j}|tweet)$ to handle underflow issue.

\subsubsection{Sentiment Analysis Results}
\begin{figure}
   \centering
   \includegraphics[width=\linewidth,height=8cm]{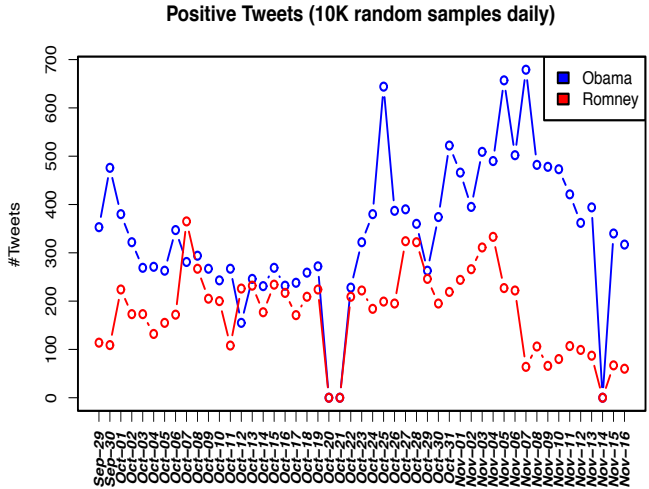}
    \caption{Obama/Romney: Positive Tweets Trend}
    \label{fig:pos}
\end{figure}

For running sentiment analysis, we randomly sampled 10K tweets from each day starting from September 29th and ending November 16th, 2012. By computing sentiment for tweets mentioning one candidate only, we can minimize the number of errors made by the NB classifier. As mentioned earlier, we cleaned each tweet by removing noise. Figure \ref{fig:pos} shows the trend of positive tweets for each candidate over the period of the election. By analyzing the number of positive tweets, we can make several observations. First, Obama led Romney in the number of positive tweets in general. Second, there is a close competition between the two candidates from October 7th until October 22th which was the day of the last presidential debate. After October 22th, we observe that Obama started leading Romney in Twitter space with a large gap.

\begin{figure}
   \centering
   \includegraphics[width=\linewidth,height=8cm]{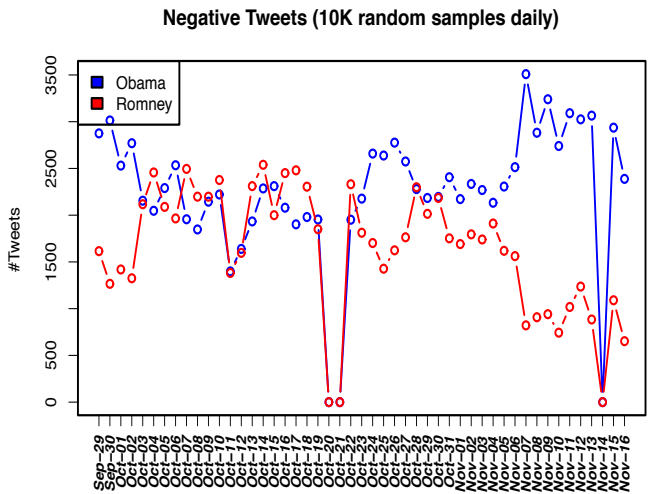}
    \caption{Obama/Romney: Negative Tweets Trend}
    \label{fig:neg}
\end{figure}

Figure \ref{fig:neg} shows the trend of negative tweets for both candidates over the election period. By analyzing the negative tweets, we make several observations. Interestingly, the number of negative tweets are significantly larger than positive ones (i.e. eight times more). This shows that in the US presidential election, both Republican and Democratic parties spent their energy to advertise negative contents for their competitor. This strategy was also dominant during the debates where Romney often attacked Obama's domestic and foreign policies. In addition, there is a close competition between the two candidates from October 3th until October 22th, 2012. Finally, after October 22th, Obama started receiving more negative tweets compared to Romney did which is due tObama's dominance in Twitter.

\subsubsection{Opinion Polls Results}
For the next analysis, we compared our Twitter sentiment analysis results with polls results collected through traditional polling systems. We accessed opinion polls data from Hffington Post website \cite{huffington:2014} where they published the polls results collected by different pollsters. Each pollster data was ran by a different poll agency where they used different tolls to interview a population sample of voters. We only considered polls data for the period of September 29th until November 5th, 2012. We found 103 number of pollster results in that time range among which 98\% of pollsters interviewed ``\textit{Likely Voters}'' and the other 2\% targeted ``\textit{Registered Voters}''. We also found that 53.4\% of polls used ``\textit{Automated Phone}'' for interviews while ``\textit{Phone}'' and ``\textit{Internet}'' portions were 24.3\% and 18.4\%, respectively. The rest was a mixture of those methods. Finally, Table \ref{tab:pollsters} shows the top pollsters in the opinion polls data.

\begin{figure}
   \centering
   \includegraphics[width=\linewidth,height=8cm]{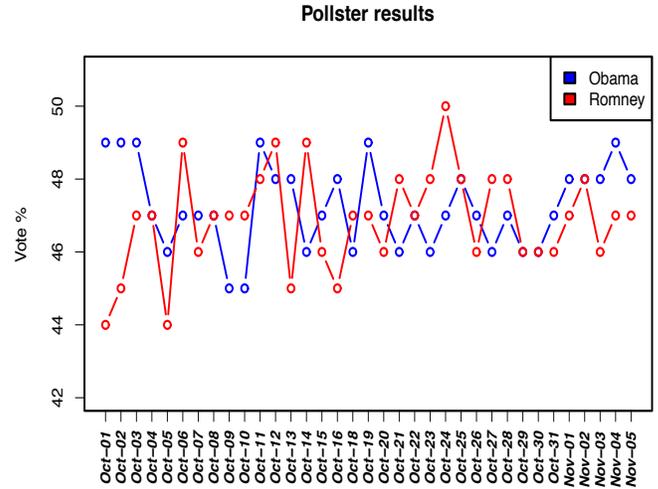}
    \caption{Obama/Romney: Polls Favorite Results}
    \label{fig:poll}
\end{figure}

Figure \ref{fig:poll} shows pollster results (i.e. favorite percentage) for each candidate before the election day. We observed that Obama and Romney had been fluctuating in polls while being behind and ahead of each other in turn from October 6th to October 29th, in a comparison of tweet mentions percentage in Figure \ref{fig:mention}, hashtags distributions on important dates in Figure \ref{fig:hashtag}, and positive tweets trend of Obama/Romney in Figure \ref{fig:pos} with pollster results in Figure \ref{fig:poll}. Moreover, according to Figure \ref{fig:poll} Obama had been ahead of Romney 17 days among 36 days from October 1st to November 5th, especially in the early October (i.e. October 1st to 5th) and the early November (November 3rd to 5th) while reaching the election day. 

The pollster results in Figure \ref{fig:poll} are mostly in match with our Twitter results but with some latency. For instance, Figure \ref{fig:mention} shows that Obama was mostly mentioned in Twitter especially from September 29th to October 3rd, and after October 23rd until November 16th (even after the election day). Obama had mostly been a popular topic in Twitter as Figure \ref{fig:hashtag} shows, and as reaching the election day, Obama was discussed about the most, more than twice compared to Romney. Finally, Obama had positive tweets much more than Romney had according to Figure \ref{fig:pos} as the election day reached. 

\begin{table}
   \centering
	\begin{tabular}{| c | c |}
	\hline
	    \textbf{Pollster} & \textbf{Frequency}\\ \hline
		Rasmussen & 12\\ \hline
		Ipsos/Reuters (Web) & 7\\ \hline
		YouGov/Economist & 6\\ \hline
		PPP (D-Americans United for Change) & 6\\ \hline
		UPI/CVOTER & 6\\ \hline
		Politico/GWU/Battleground & 6\\ \hline
		DailyKos/SEIU/PPP (D) & 5\\ \hline
		ABC/Post & 5\\ \hline
		Gallup & 5\\ \hline
		IBD/TIPP & 5\\ \hline
		ARG & 5\\ \hline
	\end{tabular}
	\caption{Frequent Pollsters}
	\label{tab:pollsters}
\end{table}

\subsubsection{Geographical Sentiment Analysis}
We have 140K tweets with geographical information (i.e. tweets posted from mobile phones) in our dataset. For a granular analysis, we repeated the sentiment analysis, focusing only on geo-tagged tweets. This spatial analysis allows us to measure the popularity of each candidate in each US state. For running geographical sentiment analysis, we stored all 50 states' latitudes/longitudes in a \textit{k-d tree} data structure \cite{michael:2002}. This data structure allows us to find the state of each geo-tagged tweet by running a nearest neighbor search in $O(\log(n))$ steps.

We define \textbf{+Ratio} to be the ratio of the number of positive tweets for Obama to the number of positive tweets for Romney in the given state. Likewise, we define \textbf{-Ratio} to be the ratio of the number of negative tweets for Obama to number of negative tweets for Romney in the corresponding state. By analyzing geo-tagged tweets for a given state, we call Obama as the winner of that state if \textbf{+Ratio} is greater than \textbf{-Ratio}. Otherwise, we call Romney the winner. Table \ref{tab:geo-sent} shows our results for a selected number of states. In Table \ref{tab:geo-sent}, \textbf{Counts} column shows the total number of processed tweets with geo-location data for the corresponding state and \textbf{Election Result} shows the final result for that state. 

In Table \ref{tab:geo-sent}, we have highlight a few states for which we could successfully predict the election results. We also show a few states for which our prediction technique failed to call the winner such as \textit{Florida}. According to our results, we could accurately predict the election results for 38 states correctly which counts for $76\%$ of states. Our predictor failed in 12 states where Obama was winner for three of those states and Romney was winner for the other nine. In the 2012 US presidential election, Obama won the election in 26 states plus DC while Romney won in 24 states. Thus, our predictor's accuracy for Obama was $85\%$ (i.e. we could predict 23 states out of 27 states including DC correctly) whereas the predictor's accuracy for Romney was only $62.5\%$ (i.e. we could predict 15 states out of 24 states correctly). Our predictor's low accuracy for Romney can be justified by Mislove et al. work in which they found that Twitter users in the US are significantly overrepresented in populous states and are underrepresented in much of mid-west states \cite{mislove:2011}.

\begin{table}
	\centering
	\begin{tabular}{ | c | c |  c | c | c | }
	    \hline
	    \textbf{State} & \textbf{+Ratio} & \textbf{-Ratio} & \textbf{Counts}  & \textbf{Election Result}\\ \hline
		\textcolor{blue}{DC} & \textcolor{blue}{1.93} & \textcolor{blue}{1.26} & \textcolor{blue}{647} & \textcolor{blue}{Obama 3}\\ \hline
		\textcolor{blue}{HI} & \textcolor{blue}{3.0} & \textcolor{blue}{1.12} & \textcolor{blue}{205} & \textcolor{blue}{Obama 4}\\ \hline
		\textcolor{blue}{CA} & \textcolor{blue}{1.69} & \textcolor{blue}{1.39} & \textcolor{blue}{7656} & \textcolor{blue}{Obama 55}\\ \hline
		\textcolor{blue}{NY} & \textcolor{blue}{1.73} & \textcolor{blue}{1.23} & \textcolor{blue}{6102} & \textcolor{blue}{Obama 29}\\	\hline
		\textcolor{red}{TX} & \textcolor{red}{1.49} & \textcolor{red}{1.56} & \textcolor{red}{7069} &\textcolor{red}{Romney 38}\\	\hline
		\textcolor{red}{AZ} & \textcolor{red}{1.14} & \textcolor{red}{3.43} & \textcolor{red}{2444} & \textcolor{red}{Romney 11}\\ \hline
		\textcolor{red}{KY} & \textcolor{red}{1.37} & \textcolor{red}{1.88} & \textcolor{red}{1593} & \textcolor{red}{Romney 8}\\	\hline
		FL & 1.120253 & 1.485866 & 8480 & Obama 29\\    \hline
	    GA & 1.865079 & 1.561066 & 3821 & Romney 16\\ \hline
		NC & 1.411765 & 1.246248 & 3621 & Romney 15\\ \hline
	\end{tabular}
	\caption{Geographical Sentiment Results}
	\label{tab:geo-sent}
\end{table}

\subsection{Analyzing Discussed Topics in Twitter}
\begin{figure}
\centering
	\includegraphics[width=7cm,height=3cm]{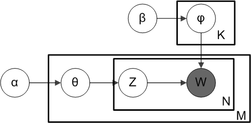}
	\caption{Graphical model of LDA algorithm}
	\label{fig:lda}
\end{figure}

As the next analysis, we processed tweets to extract the underlying topics discussed by Twitter users on each day before the election. We ran the Latent Dirichlet Allocation (LDA) model to compute the discussed topics from a large collection of tweets \cite{blei:2003}. The LDA is a general probabilistic framework for modeling sparse vectors of count data such as bags of words for text. The main idea behind the LDA algorithm for text data is that the words in each document are generated by a mixture of topics. A topic is represented as a multinomial probability distribution over words. Word distribution of each topic and topic distribution of each document are unobserved and are learned from data using unsupervised learning algorithm. Figure \ref{fig:lda} shows the graphical model for the LDA model. 

We implemented the LDA algorithm using Gibbs sampling method. We analyzed the tweets from three presidential debates: October 3rd, 16th and 22th in order to determine topics people discussed during the debate days. We set the number of topics to five ($K=5$) and ran 1000 iterations of Gibbs sampling using our ML/NLP engine. For each debate day, we sampled 10K random tweets for running the LDA model. We cleaned tweets contents using a similar technique described in the previous section. For each topic, we generated the top 15 most likely words. Tables \ref{tab:tm1}, \ref{tab:tm2}, and \ref{tab:tm3} show extracted topics from the three presidential debates.

From Table \ref{tab:tm1}, we observe that after the first debate there was a big conversation around both candidates and their performance during the debate. As expected, the word \textit{debate} was appeared in majority of topics as a word with high likelihood. This clearly shows the impact of debate on Twitter conversations. We see that \textit{tax} and \textit{middle class} also appeared as other popular topics on that day. This is because the focus of the first debate was on economy, job creation, and the federal deficit topics. Interestingly, we observe that there iwas a topic around \textit{Michelle} and \textit{anniversary} words. This is because Barack and Michelle Obama anniversary is on October 3rd, which was mentioned by both candidates in the beginning of the first debate.
 
The focus of the second presidential debate was domestic affairs and foreign policy. As we see from Table \ref{tab:tm2}, this debate started a discussion on Twitter around topics such as \textit{cut taxes}, \textit{jobs}, \textit{deficit} and the attack on the US consulate in Benghazi Libya. It is also interesting to see that the word \textit{women} appeared in one of topics. This can be because Romney used the term \textit{``Binders full of women''} when he was asked to respond to a question about pay equity for women. Thanks to Romney's comment within minutes, the hashtag \textit{\#bindersfullofwomen} was trending worldwide on Twitter. 

Finally, the third debate was focused on the foreign policy. As we see in Table \ref{tab:tm3}, there were discussions around \textit{foreign policy}, \textit{Iran}, and \textit{Bin Laden} topics in Twitter. Our results show that by using the LDA algorithm, we can effectively extract popular topics from thousands of tweets discussed in Twitter. Thus, we can use the LDA model in order to get insights from discussions in Twitter and to find out the subjects that matter to the public the most.

\begin{table}
	\centering
	\begin{tabular}{ | c | c | c | c | c | }
	    \hline
		$T_{1}$ & $T_{2}$ & $T_{3}$ & $T_{4}$ & $T_{5}$\\\hline
		romney &	obama &	obama	& obama & romney\\\hline
		obama	& romney &	romney &	romney &	obama\\\hline
		mitt	& \textbf{debate}	& mitt	& not	& mitt\\\hline
		\textbf{debate} &	mitt &	barack	& like	& \textbf{class}\\\hline
		poll	& tonight	& \textbf{debate}	& mitt	& like\\\hline
		election	& gt	& president &	vote	& \textbf{middle}\\\hline
		won	& presidential	& \textbf{michelle}	& don	& \textbf{tax}\\\hline
		cnn	& president &	not	& people	& not\\\hline
		amp	& vote	& amp	& voting &	\textbf{cut}\\\hline
		president	& election	& video	& amp	& says\\\hline
		street	& first	& \textbf{anniversary}	& election &	\textbf{taxes}\\\hline
		wins	& get	& would &	know	& president\\\hline
		y	& watch &	election	& \textbf{debate} &	fucked\\\hline
		de &	amp	& party	& president &	\textbf{debate}\\\hline
		 & like	& tonight	& get	& big\\\hline
	\end{tabular}
	\caption{First presidential debate}
	\label{tab:tm1}
\end{table}

\begin{table}
	\centering
	\begin{tabular}{ | c | c | c | c | c | }
	    \hline
	    $T_{1}$ & $T_{2}$ & $T_{3}$ & $T_{4}$ & $T_{5}$\\\hline
		obama	& obama	& romney &	romney & 	obama\\\hline
		romney	 & romney	& obama	& obama	& romney\\\hline
		\textbf{debate}	& \textbf{debate} &	\textbf{tax}	& mitt	& mitt\\\hline
		mitt	& president	& mitt	& like	& not\\\hline
		tonight &	gt	& \textbf{plan}	& not	& like\\\hline
		president	& ryan	& \textbf{debate}	& \textbf{debate} &	\textbf{jobs}\\\hline
		amp	& paul	& \textbf{trillion}	& \textbf{women}	& candy\\\hline
		not	& poll	& details	& get	& crowley\\\hline
		would	& won	& president	& vote	& right\\\hline
		\textbf{plan}	& mitt	& don	& president	& said\\\hline
		back	& amp	& like	& election	& president\\\hline
		one	& election	 & \textbf{cut}	& tonight	& \textbf{debate}\\\hline
		point &	tonight	 & \textbf{taxes}	& don	& amp\\\hline
		last	& voters	& amp	& amp	& \textbf{libya}\\\hline
		win	& vote	&	know	& get\\\hline
	\end{tabular}
	\caption{Second presidential debate}
	\label{tab:tm2}
\end{table}

\begin{table}
	\centering
	\begin{tabular}{ | c | c | c | c | c | }
	    \hline
	    $T_{1}$ & $T_{2}$ & $T_{3}$ & $T_{4}$ & $T_{5}$\\\hline
		romney	 & obama	& obama	& romney	& romney\\\hline
		obama	& romney	& romney	& obama	& mitt\\\hline
		\textbf{debate}	& \textbf{bin}	& \textbf{debate}	& mitt	& obama\\\hline
		mitt	& \textbf{laden} &	won &	not &	gt\\\hline
		\textbf{foreign}	& mitt	& poll	& president	& \textbf{plan}\\\hline
		\textbf{policy}	& bayonets	& president	& like	& like\\\hline
		president	& horses	& cbs	& amp	& point\\\hline
		not	& debate	& think	& get	& let\\\hline
		said	& said	& mitt	& vote	& \textbf{policy}\\\hline
		presidential	& won	& news	& don	& \textbf{foreign}\\\hline
		barack	& fewer &	cnn	& people	& romneys\\\hline
		like	& \textbf{navy}	& tie &	know	& five\\\hline
		\textbf{iran}	& say	& wins	& election &	left\\\hline
		won	& fact	& barack	& tonight	 & president\\\hline
		final	& like	& vote	& one	& take\\\hline
	\end{tabular}
	\caption{Third presidential debate}
	\label{tab:tm3}
\end{table}

\section{Conclusion}
\label{section:conclusion}
Social media has not been studied systematically enough yet. Regarding to the possibility of predicting future events using social media data, advanced machine learning techniques can help for more accurate analysis. Through analyzing social media data such as tweets, we can find interesting trends that can lead to better understanding, interpretation and insights. Although predicting future events by mining social media data is challenging, given its predictive functions, the data is appealing to be used as an important source of information for opinion mining.

Our methods for Twitter content analysis are systematically developed, so that we contend that our results are credible enough. Our results show that Obama was leading in Twitter for the 2012 US presidential election, which is in match with the outcome of the election. We also found that the negative advertising played an important role in Twitter conversations during the US presidential election. Our geographical sentiment analysis results show that geo-tweets can uncover candidates popularities in the US states. Our results also demonstrate that LDA is a powerful unsupervised algorithm. Every topic extracted by LDA defines a probability distribution over vocabulary words, through which we could extract hidden topical patterns from a large number of political tweets, and through which we could infer the underlying topic structure in Twitter.

Given certain claims on some drawbacks of social media data, it is interesting to know that social media platforms such as Twitter and Facebook and numerous others have been treated as an indispensable source for marketing, advertising, and promotions \cite{hanna:2011} as well as for detecting social trends and predicting future events. In addition, the phenomenon of the growth of companies in social media analysis implies the increasing demands for social media analysis. In facing the era of social media, social media data is an important source of information for different types of content analysis. For a further study, more improved methods in mining social media data will contribute to a better understanding of the social media landscape in more details.


\section*{Acknowledgment}
The authors would like to thank Li Yu for his contribution in building the analytics software for mining tweets and Mohammad Hajiabadi for his valuable comments to improve the quality of the paper.



%
\bibliography{report}
\bibliographystyle{plain}
%
%

\end{document}